\documentclass[twocolumn]
{aastex631}
\usepackage{graphicx}
\usepackage[normalem]{ulem}
\usepackage{amsmath}
\usepackage{dsfont}
\usepackage{subfigure}

\def\cos{\text{cos}}
\def\sin{\text{sin}}

\def\t{\text}

\usepackage{color}
\usepackage{float}
\usepackage{tensor}
\usepackage{xspace}
\usepackage{orcidlink}
\usepackage{url}
\usepackage[multiple]{footmisc}

\def\Mc{M_{\rm cloud}} 
\def\Rc{R_{\rm cloud}} 
\def\rhoc{\rho_{\rm cloud}}
\def\tff{t_{\rm ff}}
\def\nocrslarge{\rm M2e4\_noCRT}
\def\fiduciallarge{\rm M2e4\_E1\_D8e25}
\def\noswlarge{\rm M2e4\_E1\_D8e25\_noSWCR}

\def\gammacr{\gamma_{\rm cr}}

\def\ecr{e_{\rm cr}} 
\def\Fecr{F_{e, \rm cr}}

\def\epscr{\epsilon_{\rm cr}}

\begin{document}

\title{Gauging the Impact of Cosmic Ray Feedback on the Stellar Initial Mass Function }

\correspondingauthor{Margot Fitz Axen}
\email{fitza003@utexas.edu}

\author[0000-0003-1252-9916]{Margot Fitz Axen}
\affiliation{National Center for Computational Sciences, Oak Ridge National Laboratory, Oak Ridge, TN 37831 USA.}

\author[0000-0001-7220-5193]{Stella S. R. Offner}
\affiliation{Department of Astronomy, The University of Texas at Austin, 2515 Speedway, Stop C1400, Austin, TX 78712-1205, USA.}

\author[0000-0003-3729-1684]{Philip F. Hopkins}
\affiliation{TAPIR, Mailcode 350-17, California Institute of Technology, Pasadena, CA 91125, USA}

\author[0000-0002-1655-5604]{Michael Y. Grudi\'{c}}
\affiliation{Flatiron Institute, Center for Computational Astrophysics, 162 5th Ave, New York, NY 10010, USA.}

\begin{abstract}
Cosmic rays (CRs) drive ionization and influence gas dynamics in molecular clouds (MCs), potentially impacting the resulting star formation outcomes. Although previous simulations of individual star formation have included methods for cosmic ray transport (CRT), none have been large enough to resolve the stellar initial mass function (IMF). We conduct numerical simulations following the collapse of a $20000 M_{\odot}$ MC and the subsequent star formation including CRT, both with and without CRs accelerated by winds from the young massive stars, and compare against a non-CRT simulation. We show that after the first massive stars form, the cavity produced by feedback is more pronounced in the CRT simulations because the external CRs are able to propagate inwards and compress the gas into higher density structures. This increases the subsequent star formation in the cloud; by the end of the simulation, the SFE in the CRT simulation including stellar wind CRs is 43 \% higher than the non-CRT simulation. The IMF is also top heavy in comparison, with a 
slope above 1 $M_{\odot}$ that is shallower by $\sim 20$ \%. These effects are also present in the simulation without wind-accelerated CRs, but they are not as pronounced; the SFE is only 16 \% higher than the non-CRT simulation, and the IMF high-mass slope is shallower by $\sim 10$ \%. 
These results may explain some of the observed top-heavy IMFs, which typically occur in high-CR environments such as the galactic center. 
\end{abstract}

\section{Introduction}

Cosmic rays (CRs) are pervasive throughout the interstellar medium (ISM) and induce a  variety of dynamical and chemical processes. In particular, they drive ionization reactions in star-forming molecular clouds (MCs), which are optically thick to UV radiation \citep{dalgarno_2006}.
The CR ionization rate (CRIR), or rate of CR ionizations per hydrogen molecule per unit time, is used to quantify the CR level in MCs and can be inferred indirectly through measurements of the resulting products of these reactions. The measured CRIR in diffuse clouds in the Milky Way varies between $\approx 10^{-17}-2 \times 10^{-16} \rm s^{-1}$, with an average value of $\sim 5-6 \times 10^{-17} \rm s^{-1}$ \citep{obolentseva_2024, indriolo_2025}.
However, observations indicate lower CR levels in higher density regions such as high-mass star-forming regions \citep{sabatini_2020} and low-mass dense cores \citep{caselli_1998, bialy_2022} and higher levels towards SNRs  \citep{ceccarelli_2011} and towards the center of the Galaxy \citep{rivilla_2022}.

Some of this variation may come from the presence of local acceleration sources \citep{ceccarelli_2011}. Supernova remnants are the primary accelerators of galactic CRs \citep{blasi_2013}, converting $\sim 10$ \% of their energy into CRs \citep{vink_2012}. Additionally, recent studies have found evidence of $\gamma$-ray emission around star clusters, indicating the presence of high-energy CRs accelerated by high-mass stellar winds and outflows \citep{ackermann_2011, yang_aharonian_2017, yang_2018, aharonian_2019, sun_2020, liu_2022, pandey_2024, parizot_2004, bykov_2020, peron_2024a, peron_2024b,pandey_2025}.  Most recently, ionization maps of star-forming regions hint that protostellar sources may also accelerate CRs \citep{ceccarelli_2014, cabedo_2023, pineda_2024}, likely accelerated by the accretion or protostellar jet shocks \citep{padovani_2016,gaches_2018}.

There is now additional support from numerical simulations which include explicit cosmic ray transport (CRT) showing that the CRIR varies with environment and effects gas properties. Simulations of galaxy formation and evolution show that CRs drive galactic outflows and suppress star formation \citep{jubelgas_2008, uhlig_2012, booth_2013, hanasz_2013, salem_2014, pakmor_2016, simpson_2016, wiener_2017, butsky_quinn_2018, jacob_2018, chan_2019, dashyan_dobois_2020, hopkins_2020, ji_2020, farcy_2022, girichidis_2022, girichidis_2023, thomas_2023}. The magnitude of these effects depends on including plasma-dependant transport parameters \citep{girichidis_2016, simpson_2016, farber_2018, holguin_2019, semenov_2021} and on the CR diffusion coefficient \citep{commercon_2019, fitzaxen_2024}. Despite this, most simulations of MCs have assumed a spatially and temporally uniform CRIR of $\approx 10^{-17}-10^{-16} \rm s^{-1}$ to calculate the heating and cooling processes  \citep{glover_clark_2012, bate_keto_2015,gatto_2017, rowan_2020, grudic_2021, guszejnov_2021, guszejnov_2022} or have neglected CR ionization altogether. 

By not including CRT, these simulations did not capture the impact of CRIR variation on the gas temperature and dynamics, and thus the impact of CRs on the resulting stellar initial mass function (IMF) 
\citep{bastian_2010, offner_2014}. The IMF was long thought to be near-universal, with a peak at $\sim 0.1-0.3 M_{\odot}$ and a power law tail at $M_* \gtrsim 1 M_{\odot}$ such that $d\rm N_*/d \rm log M_* \propto  M^{\alpha}$, where $\alpha \sim -1.35$ \citep{salpeter_1955, kroupa_2002}. However, resolved star counts of Milky Way clusters with the Gaia mission \citep{luhman_2018, sollima_2019, mor_2019}, and in the disks of M31 and M33 with the Hubble Space Telescope \citep{weisz_2015, wainer_2024} have suggested moderate variations in $\alpha$, which may depend on the distribution of properties such as metallicity \citep{li_2023}. Recent observations also suggest larger variations in more extreme environments. Studies that probe starburst conditions such as those in high redshift galaxies \citep{cameron_2023, guo_2024} and the most massive, dense local group clusters \citep{bartko_2010, lim_2013, pang_2013, lu_2013, schneider_2018, hosek_2019} suggest a relatively top-heavy IMF. In contrast, massive early type galaxies (ETGs) show evidence of a bottom heavy IMF, with a steeper slope below 1 $M_{\odot}$ \citep{smith_2020}. 

By affecting gas temperatures, ionization, and kinematics, CRs may have a significant impact on the star formation in MCs and the resulting IMF. This is especialy true in starburst conditions where the CR energy density may be  $\sim 10^3 - 10^4$ times higher then in the Milky Way, raising the minimum gas temperature from $\sim 10$ K to 50-100 K \citep{papadopoulos_2011}. However, the CR flux may be strongly attenuated from the galactic background level in many regions of the cloud due to energy losses \citep{fitzaxen_2024}. Observations show tentative evidence that the CRIR declines in higher column density gas, which is expected from collisional losses \citep{padovani_2023}. In addition, CRs are expected to lose a significant amount of energy in MCs to the streaming instability \citep{kulsrud_pearce_1969, skilling_strong_1976, cesarsky_volk_1978}. Unlike the diffuse ionized ISM where CRs scatter off resonant extrinsic turbulence \citep{xu_2017}, CR transport in MCs is dominated by streaming along magnetic field lines at the ion-Alfv\'{e}n speed \citep{krumholz_2020, hopkins_2022b}. As the CRs stream they lose energy by generating Alfv\'{e}n waves, which are quickly thermalized and heat the gas \citep{hopkins_2022b}. This instability may be the dominant form of CR energy losses in MCs, especially for $\sim$ GeV CRs, which represent the peak of the CR energy spectrum and constitute most of the CR population. Numerical simulations have confirmed that the streaming instability has the potential to strongly limit the CR flux in MCs \citep{everett_zweibel_2011, bustard_zweibel_2021, fitzaxen_2024}. 

In this study we extend the work of \cite{fitzaxen_2024}, who ran simulations following the collapse of a 2000 $M_{\odot}$ cloud that included CRT and resolved the formation of individual stars using the STARFORGE (STAR FORmation in Gaseous Environments) framework for star formation \citep{grudic_2021}, which is built on the {\small GIZMO} simulation code \citep{hopkins_2015}. Those clouds only formed $\sim 120-170$ stars each, which was not enough to accurately resolve the IMF. Here, we use the STARFORGE framework to model the evolution of 20000 $M_{\odot}$ clouds to investigate the impact of CRT and stellar wind CR feedback on the IMF. Section \ref{section:methods} describes the physics and numerical methods used. Section \ref{section:results} presents the results of our simulations. Section \ref{section:discussion} summarizes our conclusions on how CRs impact the IMF.

\section{Methods}
\label{section:methods}

\subsection{The STARFORGE Simulations}
\label{subsection:starforge_simulations}

\subsubsection{Physics}
\label{subsubsection:starforge_physics}

We simulate star-forming MCs using the STARFORGE framework \citep{grudic_2021}, which uses the {\small GIZMO} simulation code \citep{hopkins_2015}. We use the Lagrangian constrained-gradient meshless finite-mass method for magnetohydrodynamics (MHD) \citep{hopkins_raives_2016, hopkins_2016} and assume the ideal MHD limit. The simulations utilize the updated FIRE-3 radiation and chemistry modules \citep{hopkins_2023}. They co-evolve the gas, dust, and radiation temperature from the stellar luminosity and an external radiation field. All quantities have periodic boundary conditions.

Sink particles represent individual stars and follow the subgrid protostellar evolution model in \cite{offner_2009}. The STARFORGE framework includes stellar feedback from protostellar jets, radiation, stellar winds, and supernovae (SNe). Feedback from SNe, radiation, and most stellar winds is injected by distributing mass, momentum, and energy into surrounding gas cells using the weighting scheme described in \cite{hopkins_2018}. Feedback from protostellar jets and stellar winds, where the free-expansion radius is smaller than the size of a wind cell is injected, use the cell spawning technique described in \cite{grudic_2021}. A brief description of the methods used for the protostellar jets and stellar winds is described in \cite{fitzaxen_2024}. 

Stellar winds are launched for stars more massive than $2 M_{\odot}$ and lose mass isotropically at a mass-loss rate $\dot{M}_w$ moving at velocity $v_w$. To test the impact of including CR acceleration from stellar wind sources, we run two simulations. The first run is identical to the prescription described in \cite{fitzaxen_2024}. Stellar winds with velocities $v_w > 1000 $ km/s inject $\eta= 10 \%$ of their energy into CRs \citep{aharonian_2019, pandey_2024} and the wind velocity is reduced to $v_w \rightarrow \sqrt{(1-\eta)}v_w = \sqrt{0.9}v_w$. We note that this choice was made to conserve total energy; however, it is not formally correct. CRs are accelerated at the wind termination shock, which is where energy dissipation occurs and shock heating is suppressed. Therefore, to first approximation it is the downstream temperature that should be modified (and not the wind speed). However, uncertainties in the stellar wind properties are larger than a reduction in the wind velocity by 10 $\%$; therefore, this choice should serve as an effective proxy to test the impact of CR injection from winds.

For the second simulation, we continue the evolution of the first simulation from $t= 1.33 \tff$ (approximately when the first massive stars $\gtrsim 10 M_{\odot}$ begin to form) with $\eta= 0$. Although this turns off stellar wind CR sources in the simulation, every stellar wind velocity is still reduced by $v_w \rightarrow \sqrt{0.9}v_w$. This choice was made to ensure that variations between the two simulations were not due to significant differences in the total wind kinetic energy; however, we do not expect this to have a strong impact on the results \citep{guszejnov_2021}. 

The STARFORGE framework  also includes SNe, which occur at the end of the lifetime of main sequence stars $> 8 M_{\odot}$. The prescription for SNe CR injection is identical to that for stellar winds, with $10\%$ of their energy lost to CR acceleration \citep{vink_2012}. While CR injection by SNe would temporarily dominate over that of stellar winds, they occur after the cloud is already dispersed by other feedback, and thus do not change the star formation outcomes of the cloud \citep{grudic_2022, guszejnov_2022}. Therefore, we do not explore the effects of these CRs in our simulations.

\subsubsection{Initial Cloud Properties}
\label{subsubsection:cloud_properties}

Our initial cloud properties are similar to those used in previous STARFORGE simulations \citep{guszejnov_2021, guszejnov_2022, grudic_2021, grudic_2022}. We start with a cloud of gas with mass $\Mc = 20000 M_{\odot}$ and radius $\Rc = 10$ pc at a uniform density of $\rhoc = \Mc/(4\pi\Rc^3/3)$ and a uniform temperature $T_0 = 10$ K. The cloud is placed in a diffuse medium with a density of $\rhoc/1000$, within a periodic box of size $10\Rc$. The initial velocity field is initialized so that the velocity power spectrum varies with wavenumber as $E_k \propto k^{-2}$. Velocities are scaled to the value set by the turbulent virial parameter $\alpha_{\rm turb} = 2$, which measures the relative importance of turbulence and gravity
\citep{bertoldi_mckee_1992, federrath_klessen_2012}.
The clouds have a uniform magnetic field $B_z$ set by the mass-to-flux ratio $\mu=1.3$ \citep{mouschovias_spitzer_1976}. We use a mass resolution of $10^{-3} M_{\odot}$ for gas cells, so each cloud initially consists of $2 \times 10^7$ gas cells. 

Table \ref{table:simulation_initial_conditions} restates the cloud properties in our simulations.  Throughout this study, we present results in terms of the cloud freefall time, which is $\tff \approx$ 3.7 Myr. We run the simulations until the clouds are completely dispersed by feedback, which occurs at $\sim $ 9.8 Myr (or 2.66 $\tff$).

\begin{table*}
\centering
\begin{tabular}{ | c c c c c c |} \hline
 $M_{\rm cloud} (M_{\odot})$ & $R_{\rm cloud}$ (pc) & $\alpha_{\rm turb}$ & $\mu$ & $T_0$ (K) & $\Mc/\Delta m$ \\
 \hline
 $2 \times 10^4$ & 10 & 2 & 1.3 & 10 & $2 \times 10^7$ \\
 \hline
\end{tabular}
\caption{Initial conditions of the simulated clouds. Columns are the initial cloud mass, size, virial parameter, mass to magnetic flux ratio, temperature, and mass resolution \citep{guszejnov_2022}.}
\label{table:simulation_initial_conditions}
\end{table*}

\subsection{Cosmic Ray Methodology}
\label{subsection:cosmic_rays}

\subsubsection{Cosmic Ray Physics}
\label{subsubsection:cr_physics}

We follow the same CR transport methods described in \cite{fitzaxen_2024}, which we describe briefly here. 

Our study extends the STARFORGE simulation suite by including the {\small GIZMO} module for ``single energy bin'' CRT, which represents the $\sim 0.5-10\,$ GeV protons containing most of the CR energy and pressure. This implementation treats CRs as a relativistic fluid ($\gammacr=4/3$) obeying an appropriate set of fluid-like equations. We evolve the total CR energy $\ecr$ and the CR energy flux $\Fecr$ with a set of moment equations using a reduced-speed-of-light (RSOL) approximation \citep{hopkins_2022a}. The RSOL approximation is equivalent to replacing $c^{-1}D_t\ecr \rightarrow \tilde{c}^{-1}D_t\ecr$ and  $c^{-1}D_t(\Fecr/c) \rightarrow \tilde{c}^{-1}D_t(\Fecr/c)$ in the original CR transport equation, and we use a value of $\tilde{c} = 300 \rm km/s$.  These equations include CR transport from advection, streaming and diffusion, CR energy and momentum injection, and CR collisional and streaming instability energy losses. The streaming velocity is approximated as the ion- Alfv\'{e}n velocity. The moment equations are closed using the \cite{levermore_1984} closure relation. 

For the CR transport, we use the value of the constant diffusion coefficient $D_{\rm FLW} = 8.33 \times 10^{25} \rm cm^2/s$ derived in Section 2.2.5 of \cite{fitzaxen_2024} following \cite{sampson_2022}. The model in \cite{sampson_2022} 
assumes that streaming dominates the CR transport and that CR diffusion is dominated by random magnetic field line walk ($D \sim D_{\rm FLW}$). We note that $D_{\rm FLW}$ is much lower than the average galactic diffusion coefficient $D_{\rm obs} \sim 10^{28} \rm cm^2/s$ \citep{evoli_2019}. Measurements of the galactic diffusion coefficient also include a component from unresolved CR streaming, which  we resolve explicitly in the CR transport equations. Additionally, in the ionized ISM where these measurements are taken, CR transport is dominated by pitch-angle scattering off of extrinsic magnetic turbulence ($D_{\rm obs} \sim D_{\rm micro} \sim D_{\rm et}$). In low-ionization environments such as MCs, ion-neutral damping prevents these turbulent fluctuations from cascading down to the scales of CR gyroradii \citep{krumholz_2020}. 

Although CRs pitch-angle scatter off of self-generated magnetic field fluctuations due to the streaming instability ($D_{\rm micro} \sim D_{\rm SI}$), \cite{sampson_2022} assume that this is subdominant compared to $D_{\rm FLW}$. To test this assumption, \cite{fitzaxen_2024} included a `variable diffusion coefficient' simulation that allowed the diffusion coefficient to vary spatially and temporally based on the plasma parameters. This simulation calculated $D_{\rm micro}$ self-consistently based on the local gas and CR properties and neglected field line wandering ($D_{\rm FLW} \rightarrow 0$). Overall, they found that computing $D_{\rm micro}$ self-consistently in this simulation typically produced \textit{lower} values for $D_{\rm micro}$ than the fiducial value of $D_{\rm FLW}$, and gave results very similar to the fiducial model. This shows that the assumptions of \cite{sampson_2022} are reasonable for our plasma conditions.

We model CR energy losses from both collisional interactions with non-relativistic particles and from Alfv\'en wave generation due to the streaming instability. To model CR collisional energy losses, we follow \cite{fitzaxen_2024} and adopt an estimate for combined hadronic, Coulomb, and ionization losses \citep{volk_1996, ensslin_1997, guo_oh_2008}. Integrating the CR flux spectrum in the solar neighborhood over all CR energies and species for the energy density and ionization rate gives a value of $\epscr \approx 1 \rm eV/cm^3$ and $\zeta \approx 1.6\times 10^{-17} \rm s^{-1}$ respectively \citep{cummings_2016, stone_2019}; therefore, {\small GIZMO} assumes a CRIR of $1.6 \times 10^{-17} \rm s^{-1}$ per $\rm eV/cm^3$ see discussion in Section \ref{subsection:caveats}). Following \cite{guo_oh_2008}, we assume 1/6 of the hadronic losses and all non-hadronic losses are thermalized \citep{mannheim_schlickheiser_1994}, adding a volumetric heating component to the gas. Streaming instability energy losses are also assumed to be quickly thermalized. 

In addition to solving equations for the CR energy and flux, the default {\small GIZMO} gas energy and momentum equations are modified to account for CR interactions with the gas. The modifications to the momentum equation arise from the CR pressure gradient doing work on the gas. The modifications to the energy equation balance the CR energy losses from the work done on the gas, streaming instability and collisional energy losses, and injection and re-acceleration energy gain. 

\subsubsection{Initial CR Configuration}
\label{subsubsection:initial_cr_configuration}

We choose our initial CR configuration for our fiducial simulation following \cite{fitzaxen_2024}. We assume an initially uniform background $\epscr, _{ \rm med}$ for the CR energy density in the diffuse medium outside the cloud. We choose our fiducial value of $\epscr, _{ \rm med} = 1 \rm eV/cm^3$ to be consistent with measurements of the CR energy density in the solar neighborhood \citep{cummings_2016, stone_2019}. Inside the cloud, we start with 10 \% of the medium CR energy density ($\epscr, _{\rm cloud} =0.1\rm eV/cm^3$) to reflect probable attenuation in the dense gas \citep{padovani_2009, padovani_2020}. However, low resolution tests show that the simulations progress similarly independently of whether the initial value is $\epscr, _{\rm cloud}=1 \rm eV/cm^3$ or $0.1 \rm eV/cm^3$ (Appendix \ref{section:cr_init_appendix}). 

\begin{table*}
\centering
\begin{tabular}{ |c c c c c| } \hline
 Simulation & $M_{\rm cloud} [M_{\odot}]$ & $\epscr,_{\rm med} [\rm eV/cm^3]$ & $D_{\rm ||} [\rm cm^2/s]$ & $\eta$  
 \\
 \hline
M2e4\_E1\_D8e25 & $2 \times 10^4$ & 1 & $8.33 \times 10^{25}$ & 0.1 \\
M2e4\_noCRT & $2 \times 10^4$ & 0 & N/A & N/A \\
M2e4\_E1\_D8e25\_noSWCR & $2 \times 10^4$ & 10 & $8.33 \times 10^{25}$ & 0 \\
 \hline
\end{tabular}
\caption{Summary of the simulation parameters. Columns are the simulation name, cloud mass, initial CR energy density (outside the cloud), CR diffusion coefficient, and stellar wind CR injection efficiency.}
\label{table:simulation_cr_initial_conditions}
\end{table*}

As discussed in Section \ref{subsubsection:starforge_physics}, we run a second simulation started from our fiducial simulation at $1.33 \tff$ without stellar wind CR feedback. These are described in Table \ref{table:simulation_cr_initial_conditions}. Run \nocrslarge\ was run prior to this study and has been used for comparison to other STARFORGE simulations. This simulation does not explicitly model CR transport but instead assumes a fixed CR energy density of $\epscr = 1 \rm eV/cm^3$ (or a CRIR of $\zeta \approx 1.7 \times 10^{-17}\,\rm s^{_1}$), which is used only to calculate the heating and cooling of the gas. 

\section{Results}
\label{section:results}

\subsection{Cloud Evolution}
\label{subsection:cloud_evolution}

\subsubsection{Overview}
\label{subsubsection:overview}

We begin with a comparison of the cloud evolution for the \fiduciallarge, \noswlarge, and \nocrslarge\ simulations. Before $\sim 1.33 \tff$, the evolution of the cloud gas for the \fiduciallarge\ and \noswlarge\ simulations follows a similar pattern to the simulations in \cite{fitzaxen_2024} and other STARFORGE simulations \citep{grudic_2022, guszejnov_2021, guszejnov_2022}. Turbulence in the cloud leads to the formation of dense filaments which contain gravitationally unstable cores. By 2.8 Myr ($\sim 0.75 \tff$), some of these cores have collapsed to form the first stars. The stars launch protostellar outflows which disrupt the gas and stir turbulence in the cloud. 

\begin{figure*}[hbt!]
\centering
\includegraphics[width=0.99 \linewidth]{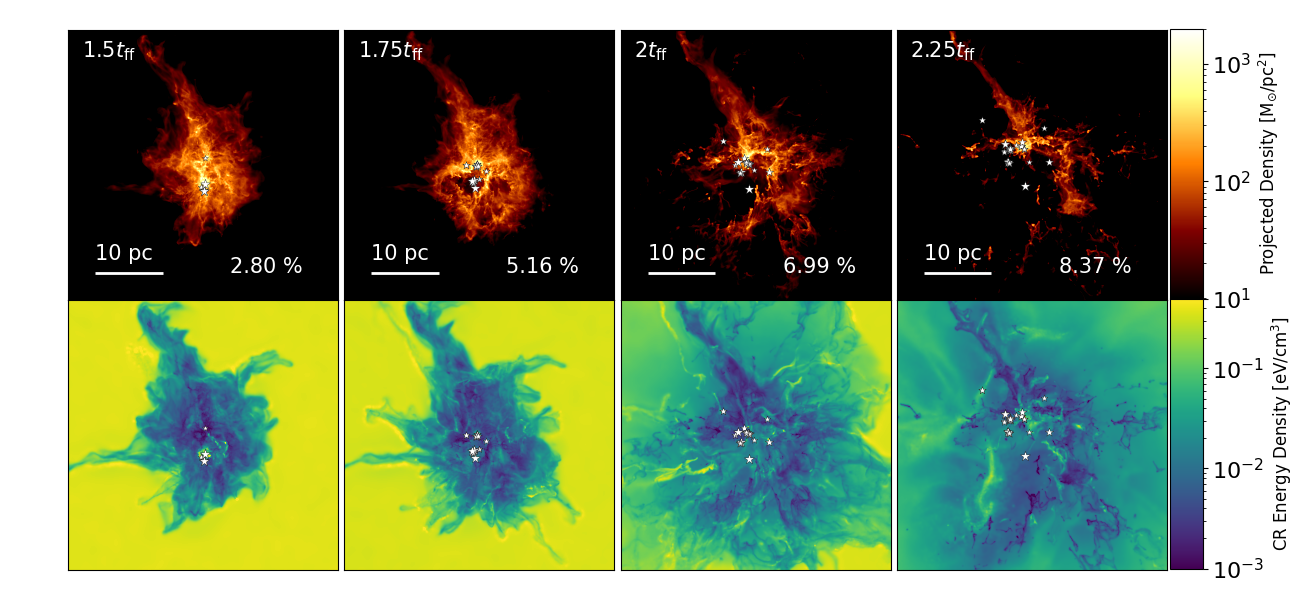}
\includegraphics[width=0.99 \linewidth]{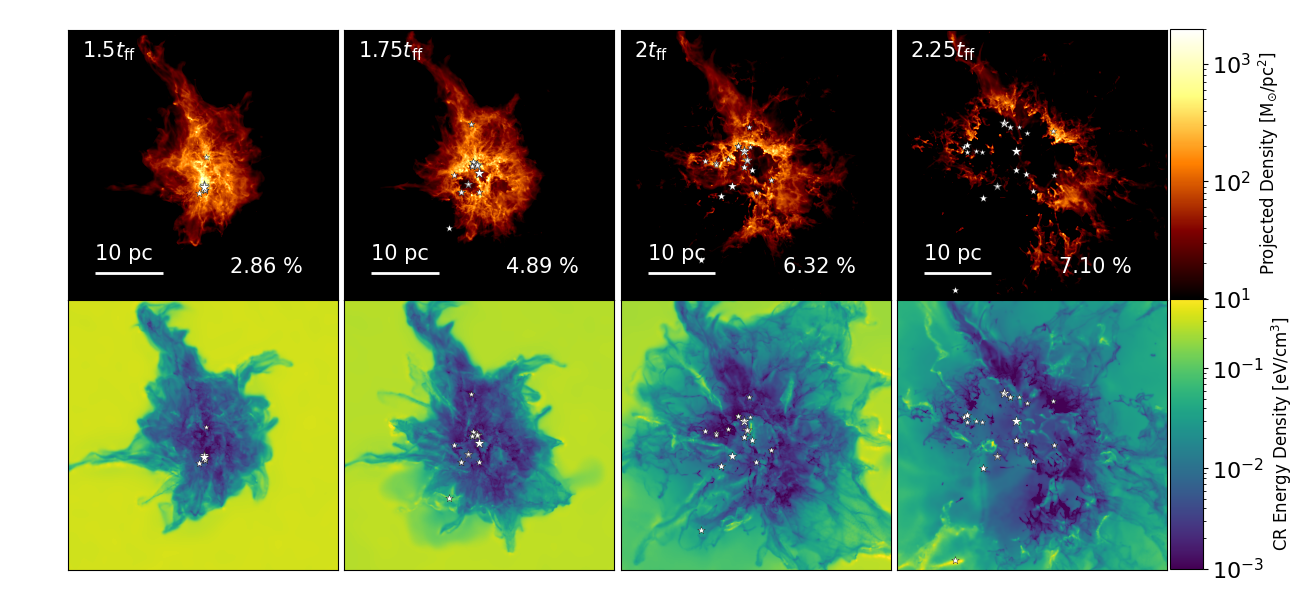}
\includegraphics[width=0.99 \linewidth]{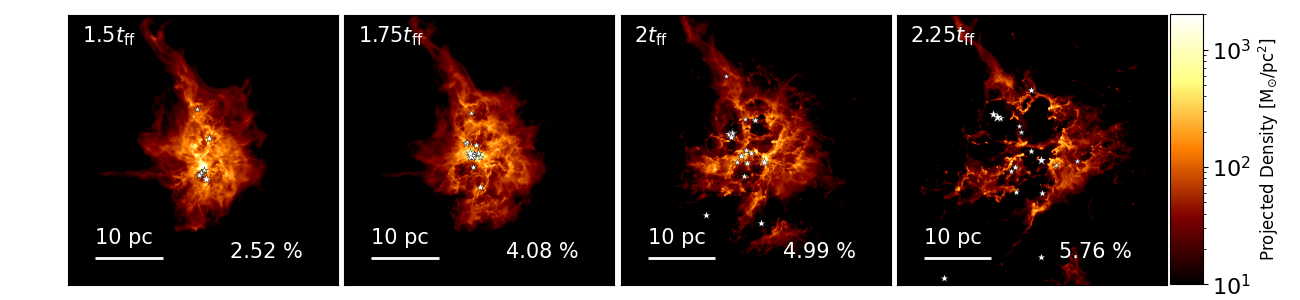}
\caption{Projected gas density and density-weighted CR energy density for the simulation with CR transport and stellar wind CR feedback, \fiduciallarge\ (first and second row) and the simulation with CR transport but no stellar wind CR feedback, \noswlarge\ (third and fourth row), and projected gas density for the simulation without CR transport \nocrslarge\ (fifth row) at four different times between $1.5t_{\rm ff}-2.25t_{\rm ff}$ ($\sim 5.6-8.3$ Myr). The SFE at each time is given in the bottom right corner. Sink particles with masses $M_* > 10 M_{\odot}$ are plotted in white, sized by their mass.}
\label{fig:proj_plots}
\end{figure*}

After $\sim 1.33 \, \tff$, when the first massive stars reach the minimum mass for stellar wind launching, the three different simulations evolve differently.  Figure \ref{fig:proj_plots} shows the time evolution of the projected gas density and density-weighted CR energy density after $1.5 \, \tff$ for the \fiduciallarge\ simulation (first and second rows) and \noswlarge\ simulation (third and fourth rows), and the projected gas density for the \nocrslarge\ simulation (bottom row) with the positions of stars $> 10 M_{\odot}$ superimposed in white. In all three simulations, massive stellar feedback creates a central cavity which causes the surrounding gas to expand and eventually disperse the cloud. However, the formation of the cavity follows a different progression between the three different simulations. By $2 \tff$ ( third column), the cavity is most noticeably pronounced in the \fiduciallarge\ simulation and least pronounced in the \nocrslarge\ simulation. 

\begin{figure}[th!]
\centering
\includegraphics[width=0.49 \textwidth]{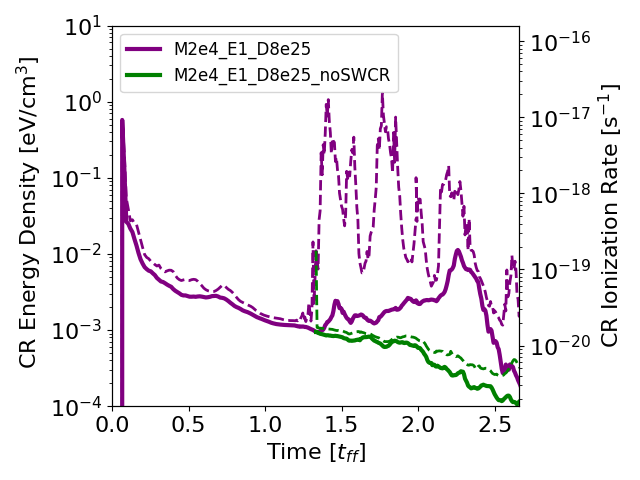}
\caption{Median (solid) and mean (dashed) CR energy density (left axis) and the corresponding CRIR assuming a solar neighborhood CR spectrum} (right axis) for gas at $n> 10^4 \rm cm^{-3}$ for the \fiduciallarge\ (purple) and \noswlarge\ (green) simulations.
\label{fig:fid_cr_energy_density}
\end{figure}

Meanwhile, at early times, the evolution of the CR energy density in the \fiduciallarge\ simulation follows the same pattern as the CRT simulations presented in \cite{fitzaxen_2024}. Inside the cloud, the CRs quickly lose energy, creating a relatively uniform CR energy density that is independent of gas density. Figure \ref{fig:fid_cr_energy_density} plots the evolution of the median (solid) and mean (dashed) CR energy density for the \fiduciallarge\ (purple) and \noswlarge\ (green) simulations at gas densities $n > 10^4 \, \rm cm^{-3}$. By $\sim 1.33 \tff$ ($\sim 4.9$ Myr), the CR energy density in the \fiduciallarge\ cloud is $\sim 10^{-3} \rm eV/cm^3$, two orders of magnitude lower then the initial cloud value of $\sim 0.1 \rm eV/cm^3$.

To understand this result, \cite{fitzaxen_2024} computed the relevant timescales for CRs to propagate through the cloud due to advection, streaming, and diffusion, and compared them to the energy loss timescales through streaming instability losses and collisional losses. They found that inside the cloud streaming transport and energy losses dominate because of the low ionization fraction in the cloud ($\chi \lesssim 10^{-7}$). CRs stream at the ion- Alfv\'{e}n velocity, which is inversely proportional to the ionization fraction ($v_{\rm st}=v_{\rm Ai}=B/\sqrt{4 \pi \chi \rho_{\rm gas}}$); thus, inside the cloud CR streaming transport and energy losses are much more efficient than outside the cloud. They also found that the streaming energy loss timescale reaches a minimum at the cloud boundary region ($n \sim 1-10 \, \rm cm^{-3}$) where the ionization fraction drops sharply. Due to this effect, CRs from outside the cloud cannot overcome the cloud boundary to repopulate the CR energy density inside the cloud.

 After $\sim 1.33 \tff$, the massive stars in the \fiduciallarge\ simulation inject CR energy and momentum as they lose mass. At this time, we begin the \noswlarge\ simulation to follow the progression of the cloud with CRT but no stellar wind CR acceleration. Figure \ref{fig:fid_cr_energy_density} shows that in the \fiduciallarge\ simulation (purple) the mean CR energy density in the dense gas increases and then varies between $\sim 1.5-2.5 \tff$. The sharp peaks and troughs are caused by different massive stars reaching the minimum mass for stellar wind launching and then rarefying the gas. This CR injection 
steadily raises the median CR energy density in the cloud as well. In the \noswlarge\ simulation (green lines in Figure \ref{fig:fid_cr_energy_density}) the mean and median CR energy density in the cloud continue to decline even after stellar winds begin. By $2 \, \tff$ the CR distribution is noticeably higher throughout the cloud in the \fiduciallarge\ than the \noswlarge\ simulation.

\begin{figure}[th!]
\centering
\includegraphics[width=0.49 \textwidth]{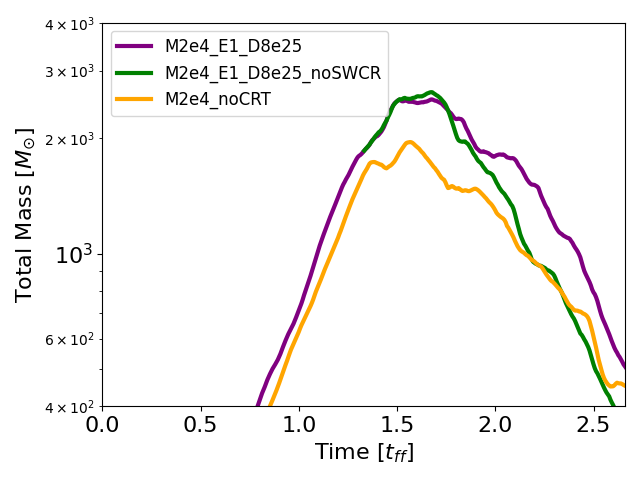}
\caption{Evolution of the total amount of gas with density $n > 10^4 \rm cm^{-3}$ for the three simulations.}
\label{fig:dense_gas}
\end{figure}

 To quantify the differences in the gas structure shown in Figure \ref{fig:proj_plots}, we plot in Figure \ref{fig:dense_gas} the evolution of the total gas mass at $n>10^4 \rm cm^{-3}$ for the three simulations. The total amount of dense gas steadily increases and peaks for all three simulations at $\sim 1.5 \tff$. Before $\sim 1.33 \tff$, the \fiduciallarge\ cloud has slightly higher gas densities than the \nocrslarge\ cloud; however, this difference becomes more noticeably pronounced after $1.33 \tff$ when massive star feedback begins. Although both clouds begin to be dispersed by feedback after $1.5 \tff$ and the amount of dense gas decreases, the \fiduciallarge\ simulation disperses more slowly than the \nocrslarge\ simulation and has more dense gas for the rest of the cloud evolution. 

Figure \ref{fig:dense_gas} shows that the \noswlarge\ (green) simulation has similar gas densities to the \fiduciallarge\ simulation until $\sim 1.85 \tff$. Although there is no stellar wind CR injection in this simulation, the evolution of the cloud is still strongly influenced by the ambient CRs. At $\sim 1.85 \tff$, the amount of dense gas in the \noswlarge\ simulation starts to decline, and by $\sim 2 \tff$ it is between the \fiduciallarge\ and \nocrslarge\ simulations.

\subsubsection{Impact of CRs on Gas Properties}
\label{subsubsection:crs_impact}

\begin{figure}[th!]
\centering
\includegraphics[width=0.49 \textwidth]{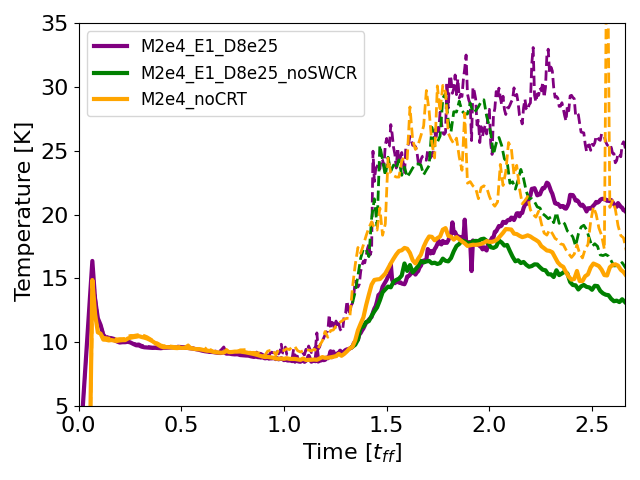}
\caption{Median (solid) and mean (dashed) temperature of the gas with $n> 10^4 \rm cm^{-3}$ for the \fiduciallarge\ (purple), \noswlarge\ (green), and \nocrslarge\ simulations.}
\label{fig:fid_temperature}
\end{figure}

Figures \ref{fig:proj_plots} and \ref{fig:dense_gas} show that including CRT changes the cloud collapse; however, these differences are not due to differences in the gas temperature. Similar to \cite{fitzaxen_2024}, we find that the effect of including CRT on the gas temperature in the cloud is minimal. Figure \ref{fig:fid_temperature} plots the evolution of the median (solid) and mean (dashed) gas temperatures at $n > 10^{4} \, \rm cm^{-3}$ for all three simulations. At early times, the gas temperature throughout the cloud is $\sim 10$ K for all three simulations. After $\sim$ 1.33 $\tff$, the temperature in the cloud increases. However, the mean temperature remains almost identical between the three simulations until $\sim 1.7 \, \tff$, which suggests that this is due to radiative heating from the stars. More significant differences in the gas temperature after $\sim 1.7 \tff$ can be ascribed to differences in the stellar mass distribution (Section \ref{subsection:star_formation_properties}).

\begin{figure*}[htp]
  \centering
  \subfigure{\includegraphics[scale=0.33]{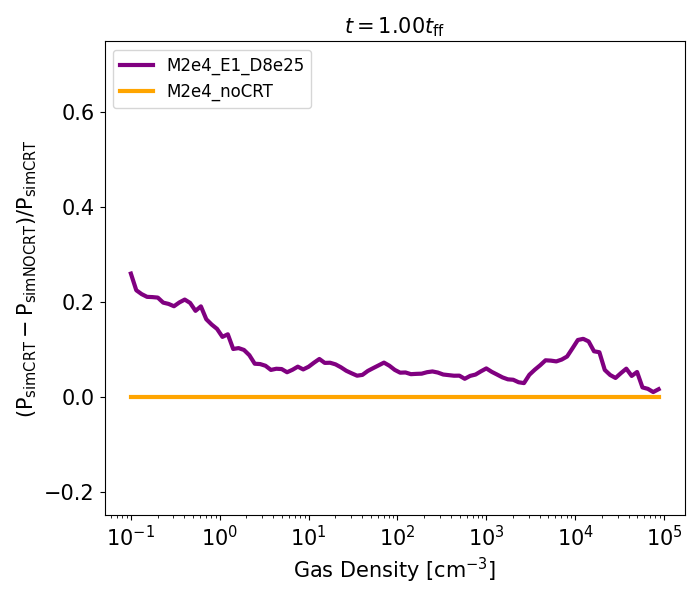}}
  \subfigure{\includegraphics[scale=0.33]{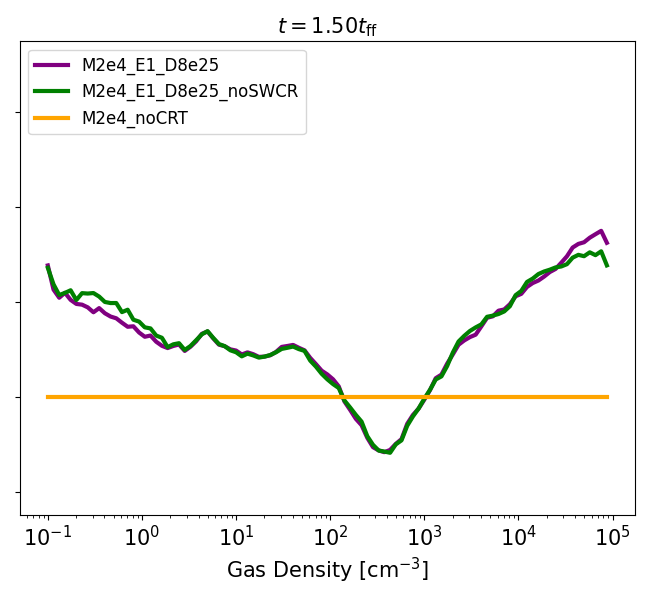}}
  \subfigure{\includegraphics[scale=0.33]{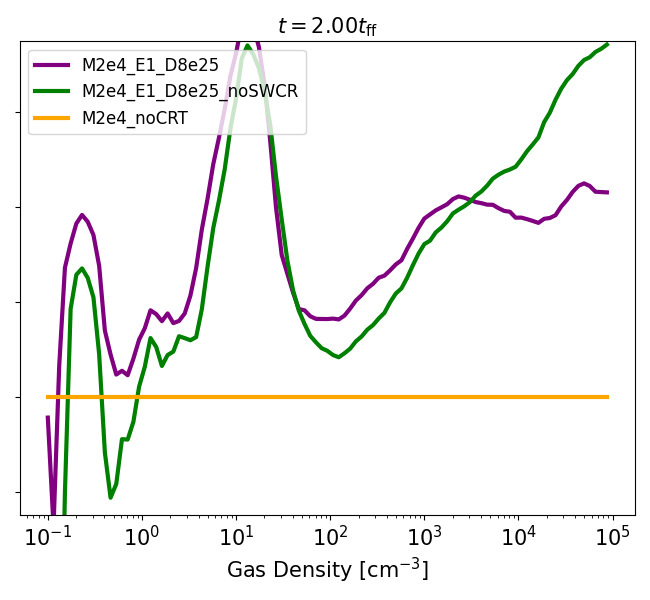}}
  \caption{Relative difference in the total gas pressure $P_{\rm tot}$ as a function of gas density between two simulations at $\tff$ (left), $1.5 \tff$ (middle),  and $2 \tff$ (right). The total pressure in one cell is calculated as $P_{\rm tot} = P_{\rm CR}+P_{\rm B}+P_{\rm K}+P_{\rm th}$, where $P_{\rm CR}$, $P_{\rm B}$, $P_{\rm K}$, and $P_{\rm th}$ are the CR, radiative, magnetic, kinetic, and thermal pressures from Equations \ref{eq:p_cr}, \ref{eq:p_b}, \ref{eq:p_k}, and \ref{eq:p_th} respectively. The lines indicate the relative pressure difference between a run with CRT (either \fiduciallarge\ or \noswlarge) and the run with no CRT (\nocrslarge).}
  \label{fig:pressure}
\end{figure*}

The CRs have a dynamical effect that changes the pressure on the gas. Figure \ref{fig:pressure} shows the relative difference in the median total pressure $P_{\rm tot}$ as a function of gas density between two simulations. The total pressure in one cell is calculated as $P_{\rm tot} = P_{\rm CR}+P_{\rm B}+P_{\rm K}+P_{\rm th}$, where $P_{\rm CR}$, $P_{\rm B}$, $P_{\rm K}$, and $P_{\rm th}$ are the CR, magnetic, kinetic, and thermal pressures respectively. These pressure contributions are calculated from the gas and CR quantities as 
\begin{equation}
P_{\rm CR} = (1/3) \epsilon_{\rm cr},
\label{eq:p_cr}
\end{equation}
\begin{equation}
P_{\rm B} = (1/8\pi) B^2,
\label{eq:p_b}
\end{equation}
\begin{equation}
P_{\rm K} = (1/2) \rho \sigma_v^2,
\label{eq:p_k}
\end{equation}
\begin{equation}
P_{\rm th} = \rho N_A k_B T, 
\label{eq:p_th}
\end{equation}
where 
$B$ is the magnitude of the magnetic field and $\sigma_v$ is the 3D velocity dispersion. Each line compares one of the two CRT simulations ($P_{\rm simCRT}$) to the \nocrslarge\ simulation ($P_{\rm simNOCRT}$). 

At early times before stellar feedback begins (left plot), the total pressure in the \fiduciallarge\ simulation is slightly higher throughout the cloud gas than the \nocrslarge\ simulation. Outside the cloud boundary ($n \lesssim 10 \, \rm cm^{-3}$) it is $\approx 15-20$ \% higher. The additional pressure comes both directly from the CRs and from additional thermal and kinetic pressure due to their changes to the gas dynamics. Figure \ref{fig:dense_gas} shows that this results in slightly earlier cloud collapse in the \fiduciallarge\ simulation; however, up to this point the cloud structure is largely the same between the two simulations (left column of Figure \ref{fig:proj_plots}). 

When stellar feedback begins, the CRs enter the cloud through the low density cavities and cause more compression of the inner cloud gas. This causes the sharp rise in dense gas mass shown in Figure \ref{fig:dense_gas} after 1.33 $\tff$ for the CRT simulations. By 1.5 $\tff$ (middle panel of Figure \ref{fig:pressure}), elevated star formation in the CRT simulations  (Section \ref{subsection:star_formation_properties}) creates more thermal, kinetic, and radiative pressure than in the \nocrslarge\ simulation, which appears at most gas densities. Differences in the star formation continue to elevate the pressure throughout the clouds in the CRT simulations, which persist until later times up until cloud dispersal (right panel) and largely remain similar between the two CRT simulations. These plots show that external CRs have a larger effect on cloud collapse than internally accelerated CRs, but only after massive stellar feedback begins when they can enter the cloud through the feedback cavities.

\subsection{Star Formation Properties}
\label{subsection:star_formation_properties}

\begin{figure*}
\centering
\includegraphics[width=0.49 \textwidth]{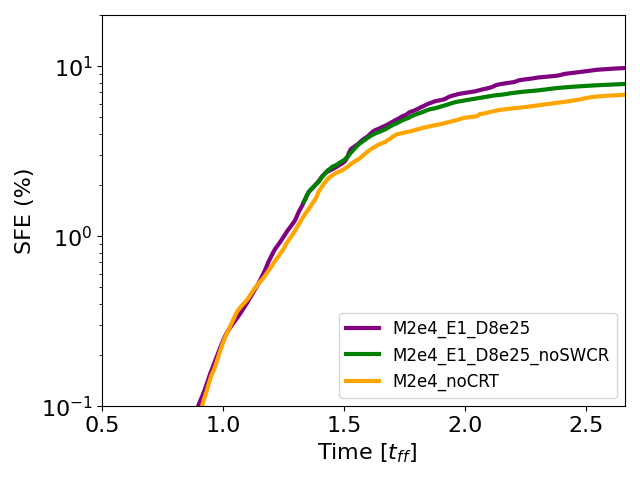}
\includegraphics[width=0.49 \textwidth]{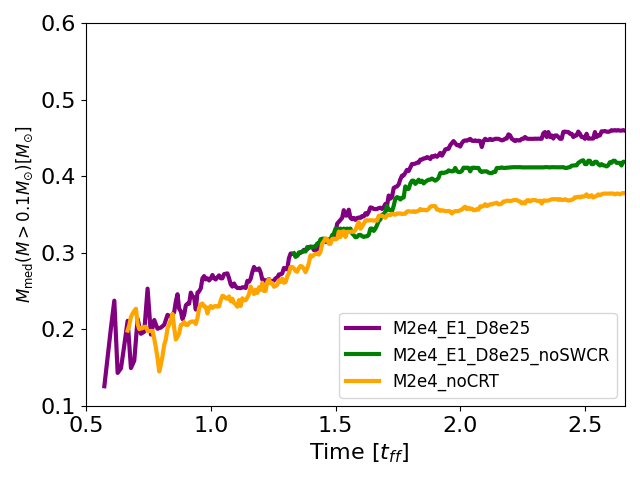}
\caption{Left: Evolution of the star formation efficiency (SFE) for all three simulations. Right: Median mass for $M_* \ge0.1 M_{\odot}$ versus time for all runs.}
\label{fig:sf_plots}
\end{figure*}

We now turn to the star formation history of the cloud. Figure \ref{fig:sf_plots}, top left, shows the evolution of the star formation efficiency (SFE), defined as the percentage of the initial cloud mass that has been converted into stars. At early times, the SFE evolves similarly for the three simulations. Once star formation starts, the SFE increases rapidly as the clouds collapse. At 1.5 $\tff$, approximately when star formation peaks, the SFE is $\sim 2.5-3 \%$ for all three simulations. After $\sim 1.5 \tff$, as feedback disperses the cloud, the slope of the SFE decreases. The growth rate of the SFE declines more rapidly for the \nocrslarge\ simulation than the \fiduciallarge\ and \noswlarge\ simulations. The second column of Table \ref{table:final_sf_parameters} shows the SFE at $2.66 \tff$ for all three simulations. By this time, the SFE of the \fiduciallarge\ and \noswlarge\ runs is 43 \% and 16 \% higher than the \nocrslarge\ run, respectively.  

The top right plot of Figure \ref{fig:sf_plots} shows the evolution of the median stellar mass above the resolution limit $0.1 M_{\odot}$. The median masses of the three simulations diverge after massive stars $\gtrsim 10 M_{\odot}$ begin to form at $\sim 1.33 \tff$. Even though feedback is dispersing the cloud and the growth of the SFE is slowing, the median mass of the \fiduciallarge\ and \noswlarge\ runs increases steeply between $\sim 1.6-1.9 \tff$. The third column of Table \ref{table:final_sf_parameters} shows the median mass at $2.66 \tff$ for all three simulations. By this time, the median masses of the \fiduciallarge\ and \noswlarge\ runs are 21 \% and 11 \% higher than the \nocrslarge\ run respectively.  

\begin{table*}
\centering
\begin{tabular}{ |c c c c c c| } 
\hline
 Simulation Name & SFE (\%) & $M_{\rm med} (M > 0.1 M_{\odot}) (M_{\odot})$ & $M_{\rm max} (M_{\odot})$ & $N_*$ & $\alpha$ ($M > 1.0 M_{\odot}$)
 \\
 \hline
M2e4\_E1\_D8e25 & 9.76 & 0.46 & 28.1 & 1471 & -0.71 $\pm$ 0.06 \\
M2e4\_E1\_D8e25\_noSWCR & 7.87 & 0.42 & 31.4 & 1328 & -0.79 $\pm$ 0.07  \\
M2e4\_noCRT & 6.81 & 0.38 & 31.2 & 1261 & -0.88 $\pm$ 0.07  \\
 \hline
\end{tabular}
\caption{Final values for the SFE, median mass above the incompleteness region, maximum mass, number of sink particles, and IMF slope above 1 $M_{\odot}$ for all simulations at 2.66 $\tff$.}
\label{table:final_sf_parameters}
\end{table*}

\begin{figure}[th!]
\centering
\includegraphics[width=0.49 \textwidth]{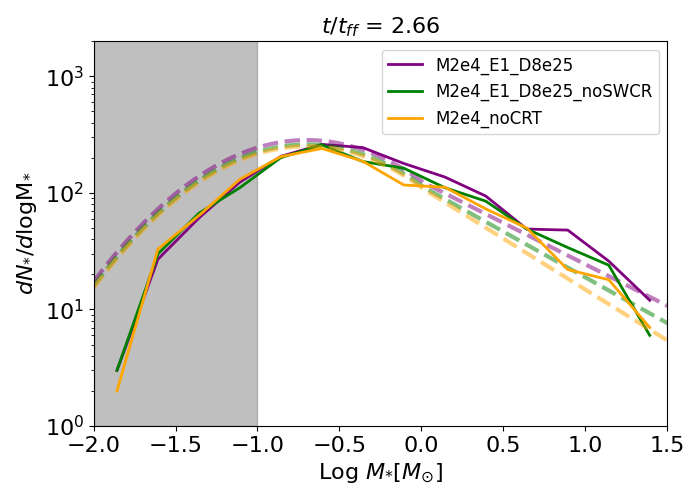}
\caption{Stellar initial mass function (IMF) for all runs at $t = 2.66\tff$. The shaded region at $M<0.1 M_{\odot}$ indicates the low-mass incompleteness region \citep{grudic_2021}. Dashed lines show a log-normal plus power-law slope function IMF \citep{chabrier_2005}, where the slope is determined by the best fit to the stellar mass distribution above 1 $M_{\odot}$ ( listed in the last column of Table \ref{table:final_sf_parameters} for each mass distribution.)}
\label{fig:imf}
\end{figure}

By 2.66 $\tff$, the three simulations have formed $\sim 1200-1500$ stars. Figure \ref{fig:imf} shows the stellar mass distribution at 2.66 $\tff$. The last column of Table \ref{table:final_sf_parameters} shows the IMF slope at $2.66 \tff$ for all three simulations.  The high mass slope of the IMF for the \nocrslarge\ simulation is shallower than the \cite{salpeter_1955} value of $\alpha=-1.35$, possibly due to missing physics or resolution (see \cite{grudic_2022} for a discussion). The elevated mass distributions in the \fiduciallarge\ and \noswlarge\ simulations result in an even shallower high mass slope.

\section{Discussion}
\label{section:discussion}

\subsection{High Mass Star Formation}
\label{subsection:high_mass_sf}

High-mass star formation is observed to occur in massive, high-column density clumps and cloud complexes \citep[][and references therein]{motte_2018}. Statistical and numerical studies suggest that the formation of high-mass stars does not simply occur through random sampling from some universal stellar mass function but instead is associated with the physical conditions characteristic of such massive regions \citep{weidner_kroupa_2006,kroupa_2021,grudic_2023}. Within the massive molecular cloud complexes imaged by the HOBYS key program, high-density dominating clumps are confirmed to be the preferred sites for forming massive stars \citep{hill_2011, nguyen_luong_2011, tige_2017}. 

 Several star formation models predict a correlation between the amount of dense gas and the stellar mass distribution, particularly the incidence of high-mass stars
\citep{mckee_tan_2003, krumholz_mckee_2008, mckee_offner_2010}. In the competitive accretion (CA) model \citep{bonnell_1997}, stars accrete gas from a common potential well until all gas is either accreted or expelled by stellar feedback.  
In this scenario the protostellar accretion rate $\dot{m}$ scales with the local freefall time and hence the gas density as $\dot{m} \propto 1/\tff \propto n^{1/2}$ \cite[e.g.,][]{bonnell_1997,bonnell_2001,mckee_offner_2010}. In the turbulent core (TC) model \citep{mckee_tan_2002}, stars form from individual gravitationally bound cores rather than within a common gas reservoir. In this model, the accretion rate scales with the surface density of the clump $\Sigma_{\rm cl}$ as $\dot{m} \propto \Sigma_{\rm cl}^{3/4}$ \citep{mckee_offner_2010}. Both of these models suggest that higher density gas leads to the formation of more high-mass stars. 

We find that increasing the amount of CRs from stellar feedback  increases the amount of dense gas and subsequent star formation. Although both CRT simulations show different evolution than the \nocrslarge\ simulation, the impact on the gas dynamics and star formation is higher for our \fiduciallarge\ simulation than our \noswlarge\ simulation. This aligns with the results 
of \cite{fitzaxen_2024}, who found that in a high CR environment ($\epsilon_{\rm cr, med} = 10 \epsilon_{\rm cr, MW}$), the cloud collapsed faster and had an elevated SFE compared to a cloud in a more typical  Milky-Way CR environment. 

In this study we did not run any simulations starting with a high CR background. However, one of the few examples of a statistically different IMF is the IMF in the galactic center, where there is an apparent excess of higher mass stars \citep{bartko_2010, lu_2013}. This environment also has a higher measured CR background \citep{rivilla_2022}. Our results and the results of \cite{fitzaxen_2024} show that the observed top heavy stellar mass distributions may be the result of both a higher CR background and internal CR acceleration. Simulations of cloud collapse in a high CR environment may show even more extreme variations in the IMF than those presented here.

\subsection{$\gamma$-ray Emission in Molecular Clouds}
\label{subsection:gamma}

Another observational constraint on the CR spectrum is provided by the production of $\gamma$-ray photons. $\gamma$-rays are produced when $\gtrsim$ 10 GeV CR protons lose their energy via pion production and where the resulting neutral pions decay into $\gamma$-rays  \citep[e.g.,][]{krumholz_2023}. \textit{Fermi}/LAT measurements of the diffuse $\gamma$-ray emission in the galactic disk suggest a relatively uniform CR energy density throughout the Galactic disk, which is consistent with small CR spatial gradients. Since MCs are a low volume filling fraction of the Galaxy, $\sim 1-2$ \% \citep{chevance_2023}, we expect that CR energy losses in MCs are offset by acceleration from SNRs and stellar winds, and thus do not significantly impact the integrated $\gamma$-ray emission from galaxies.

Other studies of the $\gamma$-ray emission in nearby GMCs with the \textit{Fermi}-LAT $\gamma$-ray telescope suggest that the flux and spectral slope of $\sim$ 10-100 GeV CRs are in good agreement with the solar neighborhood value, while the flux of $\lesssim$ GeV CRs is 
suppressed by a factor of $\sim$2 on scales of $\lesssim$ 1 pc  due to slower transport in the dense gas \citep{yang_2014, yang_2023}. Since we do not model a full multi-bin CR spectrum, we cannot directly compute the $\gamma$-ray emission from our simulations. In general we expect that a flatter CR spectrum, with relatively more high-energy CRs, would produce more $\gamma$-ray emission. In contrast, a reduction in high-energy CRs by a factor of 10-100 within the cloud, as shown in Figure \ref{fig:fid_cr_energy_density} for $t \lesssim 1 t_{\rm ff}$, would correspond to a reduction by a similar factor in the  $\gamma$-ray emission \citep{habegger_2025}. 

\begin{figure*}[hbt!]
\centering
\includegraphics[width=0.9 \linewidth]{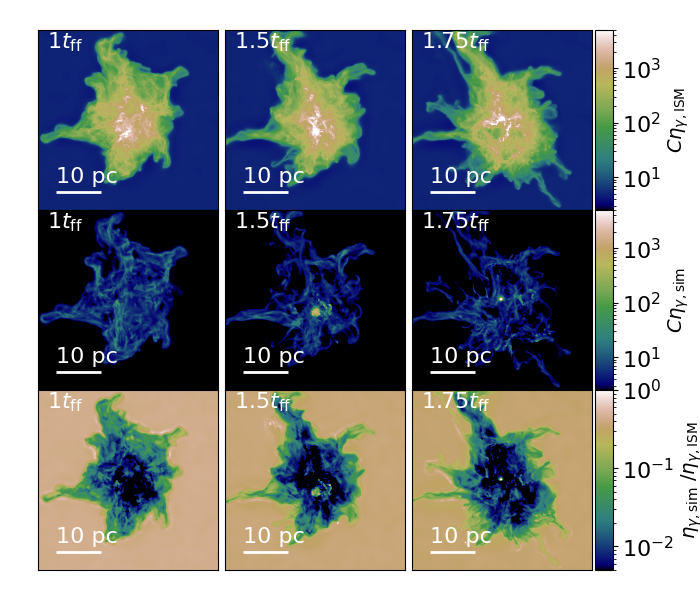}
\caption{Projected $\gamma$-ray emissivity assuming a constant CR energy density of 1 $\rm eV/cm^3$ (first row) and using the CR energy density from our simulations (second row) (both scaled by the same constant $C$ for plotting purposes), and the ratio between the two (third row) for our M2e4\_E1\_D8e25 simulation at 1.0, 1.5 and 1.75 $t_{\rm ff}$.}
\label{fig:gammas}
\end{figure*}

To investigate the potential impact of the CR attenuation on 
$\gamma$-ray production, we present maps of the $\gamma$-ray emissivity for  our M2e4\_E1\_D8e25 simulation. Since the $\gamma$-ray emissivity $\eta_{\gamma} \propto \epsilon_{\rm CR} n_H$, we compute $\eta_{\gamma, \rm sim} = \epsilon_{\rm CR}n_H$ using $\epsilon_{\rm CR}$ from the simulation and compare to $\eta_{\gamma, \rm ISM}$, which assumes  $\epsilon_{\rm CR}=1 \rm eV/cm^3$ throughout the domain. Figure \ref{fig:gammas} shows the projected 
$\eta_{\gamma, \rm ISM}$ (first row) and $\eta_{\gamma, \rm sim}$ (second row), and the ratio $\eta_{\gamma, \rm sim} / \eta_{\gamma, \rm ISM}$ (third row) at 1, 1.5, and 1.75 $t_{\rm ff}$. As expected, at all three times our clouds show lower values of the $\gamma$-ray emissivity throughout most of the domain than  for a 
uniform 1 $\rm eV/cm^3$. The CR energy density
declines by up to two orders of magnitude in the cloud center. This is 
larger than that suggested by observations. However, once stellar wind feedback begins ($\sim 1.5 t_{\rm ff}$ and later), there are hotspots in the cloud with higher $\gamma$-ray emission. Discrepancies with observations may point to the importance of CR acceleration from other star-formation sources, including jets and HII regions \citep{padovani_2016, padovani_2019}, or to a higher external CR flux than we assume here. Synthetic observations, which take into account resolution effects, are required for a detailed comparison.   

Since the resolution of \textit{Fermi}-LAT is comparable to the size of nearby GMCs and blends the emission from the cloud and its environment, results such as those from \cite{yang_2014} can not clearly distinguish between a uniform CR background and a spatially declining CR energy density that is augmented by local CR acceleration due to massive stellar feedback. These observational studies also include higher mass clouds ($\sim 10^5 M_{\odot}$), which likely contain and/or are near more massive stars producing more CR feedback (and thus more $\gamma$-ray emission) than in our simulations.  Future work, including more detailed observational modeling  to estimate the $\gamma$-ray emission \citep[e.g.,][]{krumholz_2022} would provide additional constraints on CRT in nearby GMCs.

\subsection{Simulation Caveats}
\label{subsection:caveats}

Our simulations make a number of approximations and assumptions that may have an impact on the cloud dynamics and star formation (see \cite{fitzaxen_2024} for a full discussion). In particular, variations in the CR physics may increase or decrease the magnitude of the effects of the CRs on our results.

We use a single-energy bin model for the CR fluid, which is most accurate for modeling $\sim$ GeV CR protons. {\small GIZMO} uses a normalization constant to scale all CR energy loss, ionization and heating terms from the local CR energy density. This constant was derived by taking fits to the local ISM CR spectra and integrating over all energies and species; thus, this is equivalent to assuming an underlying CR spectrum that is the same shape as that measured by the \textit{Voyager} spacecraft. However, this is not necessarily accurate, as the CR spectrum in a molecular cloud may be a different shape than the \textit{Voyager} spectrum.

Simulations modeling a full multi-bin CR spectrum and other CR species may produce different ionization in the cloud. For example, if the CR spectrum were steeper than the \textit{Voyager} spectrum, with a  higher proportion of low-energy CRs ($\lesssim 0.5$ GeV) compared to high-energy CRs ($\gtrsim 0.5$ GeV), the simulations would
underestimate the CR ionization rate for the same CR energy density, since $\sim$ MeV CRs more efficiently ionize gas than $\sim$ GeV CRs. Similarly, simulations modeling CR electrons may produce different ionization if the proportion of CR electrons to protons is different than in the local ISM. However, we do not expect either low energy CR protons or CR electrons to have a significant impact on the dynamics of cloud collapse and star formation. Moreover, we expect that both would still lose energy quickly and reach a low CR energy density throughout the cloud, even if there is more appreciable ionization along the way.

Additionally, in this study we only run simulations for one value of the CR diffusion coefficient, which is highly uncertain. The value of the diffusion coefficient impacts CR transport and energy losses in the cloud. \cite{fitzaxen_2024} showed that increasing the value of the diffusion coefficient by an order of magnitude  produces a roughly uniform CR energy density throughout the domain (both inside and outside the cloud), which is almost as high as the starting value outside the cloud ($\epsilon_{\rm CR} = 1 \rm eV/cm^3$). The CRs scatter less, and so propagate more efficiently. However, this scenario was only realized by ad-hoc setting the diffusion coefficient to a high value that is inconsistent with conditions expected in mostly neutral gas \citep{sampson_2022}.

In contrast, we expect the diffusion coefficient in the ambient medium, which has lower densities and higher temperatures, to be higher than the value of $D_{\parallel, \rm FLW}$ that we use, which was adopted to reflect physical conditions inside the cloud. Since the CRs scatter more then they should outside the cloud, streaming instability energy losses here are overestimated. By $1 t_{\rm ff}$, the median CR energy density outside the cloud has declined to $\lesssim 0.5 \rm eV/cm^3$, a reduction of $\gtrsim 50 \%$. However, we expect these losses to have a minor effect on the CR energy density inside the cloud. Since the CR streaming energy loss timescale drops sharply at the cloud boundary region, CRs from outside the cloud cannot repopulate the CR energy density inside the cloud (Section \ref{subsubsection:overview}). Thus, a marginally higher CR energy density outside the cloud would not significantly impact the CR energy density inside the cloud, as CRs cannot overcome the cloud boundary region without losing most of their energy.

Finally, the present study only includes internal CR acceleration from stellar winds with one value of the stellar wind CR injection efficiency. Our choice of $\eta=10$ \% is a common choice but is an upper limit on observational results, and the magnitude of the effects of stellar-wind CRs may be lower than those presented here. However, simulations and observations suggest that there may also be CR acceleration from protostellar jet and accretion shocks \citep{ceccarelli_2014, padovani_2016, gaches_2018, fitzaxen_2021, cabedo_2023, pineda_2024}, as well as from explosive dispersal outflows \citep{pandey_2025}, which we do not include in this study. Simulations including these CR sources will likely have a higher CR ionization rate than those presented here, particularly early in the simulation before stellar winds begin.

While our results suggest that CRs have a meaningful impact on the IMF, we caution that their statistical significance is limited given that we only run one simulation for each CR setup. Previous simulations suggest that there is significant scatter in cloud and star formation properties due to random sampling \citep{grudic_2023}, and more simulations with CRT are necessary to confirm our conclusions. Moreover, variations in other physics such as the heating and cooling prescription, cloud properties, feedback prescription, and non-ideal MHD effects also impact the slope of the IMF \citep{guszejnov_2022, grudic_inprep}, which may be larger than the effects produced by variations in the CR transport and feedback. Therefore, observations of top-heavy IMFs such as those observed in the Galactic center cannot definitively be ascribed to CRs but may instead have a number of other factors at play.

\section{Conclusions}
\label{section:conclusions}

In this work, we present numerical simulations of collapsing MCs including CRT using the STARFORGE framework. We extend the work of \cite{fitzaxen_2024} by modeling 20000 $M_{\odot}$ clouds and varying the stellar wind CR acceleration prescription to investigate the impact of CRT and stellar wind CR feedback on the IMF. We find that at early times, before massive stellar feedback begins, the CR energy density throughout the cloud is strongly attenuated due to streaming instability energy losses (Figure \ref{fig:fid_cr_energy_density}). Consequently, including CRT does not significantly change the rate of
collapse and star formation (first column of Figure \ref{fig:proj_plots}). 

After the first massive stars form, the chosen CRT prescription impacts the evolution of the cloud and star formation properties. In CRT simulations with and without stellar wind CR feedback, the cavity formed by feedback from massive stars is more pronounced at later times than in simulations without CRT (last column of Figure \ref{fig:proj_plots}). However, the gas is not more efficiently dispersed but instead compressed into higher density gas structures (Figure \ref{fig:dense_gas}). This produces differences in the star formation properties; eventually, both the SFE and median stellar mass in the clouds with CRT are higher than in the simulation without CRT (Figure \ref{fig:sf_plots}).  

By the end of the simulation, the SFE and median stellar mass in the simulation without stellar wind CR feedback are still higher than in the simulation without CRT, although both quantities are less elevated compared to the simulation with stellar wind CRs. This is because the ambient CRs are able to more efficiently propagate into the cloud through the cavity created by the massive stellar feedback and impact the ongoing star formation. The final SFEs in the clouds are 9.76 \%, 7.87 \%, and 6.81 \% for the CRT simulations with and without stellar wind CRs and the non-CRT simulation respectively. The final IMFs of the CRT simulations are top heavy compared to the non-CRT simulation, with high mass slopes of $\alpha=-0.71 \pm 0.06$ and $\alpha=-0.79 \pm 0.07$, while the non-CRT simulation shows $\alpha=-0.88 \pm 0.07$. (Figure \ref{fig:imf}). 

In summary, our results suggest that CRT and CR feedback modeling are crucial to accurately study the IMF, particularly in environments that may experience significant CR fluxes. We find that including CR transport and feedback does not necessarily decrease the star formation in the cloud, unlike the typical impact of other stellar feedback sources. Future numerical studies are needed to explore the impact of a multi-bin CR spectrum on star formation and the IMF, along with variations in the CR diffusion coefficient. Finally, simulations are needed to study both the impact of a high-CR environment and internal CR acceleration on the IMF, as the combination of these two effects may cause even more extreme variations in the IMF than are shown in the present study.

\begin{acknowledgements}
This research is part of the Frontera computing project at the Texas Advanced Computing Center, and used computing award AST21002. Frontera is made possible by National Science Foundation award OAC-1818253. This research was supported in part by NASA ATP grant 80NSSC20K0507 and NSF AAG grant 2107340. This research used resources of the Oak Ridge Leadership Computing Facility at the Oak Ridge National Laboratory, which is supported by the Office of Science of the U.S. Department of Energy under Contract No. DE-AC05-00OR22725.
\end{acknowledgements}

\appendix

\section{Choice of the Initial CR Energy Density Configuration}
\label{section:cr_init_appendix}

For this study, we use the same initial CR configuration as the fiducial model in \cite{fitzaxen_2024}. We start with $\epsilon_{\rm cr, med}=1 \rm eV/cm^3$ and $\epsilon_{\rm cr, cloud}=0.1 \rm eV/cm^3$ to reflect observations and analytic models suggesting probable attenuation in the dense gas prior to the start of the simulation \citep{padovani_2009, padovani_2023}. However, there are uncertainties in both the data and models; therefore, we test the impact of this choice by performing low resolution simulations varying the initial CR energy density in the cloud. These simulations were run with a cloud mass of $M_0=2000 M_{\odot}$ and $N_{\rm gas} = 2 \times 10^5$ (10 x lower mass resolution than our standard simulations). One of these simulations begins with a setup identical to that of our fiducial simulation ($\epsilon_{\rm cr, cloud}$=0.1 $\epsilon_{\rm cr, med}$ = 0.1 $\rm eV/cm^3$) and one starts with the initial CR energy density in the cloud equal to that of the medium ($\epsilon_{\rm cr, cloud}$= $\epsilon_{\rm cr, med}$ = 1 $\rm eV/cm^3$).

\begin{figure}[th!]
\centering
\includegraphics[width=0.49 \textwidth]{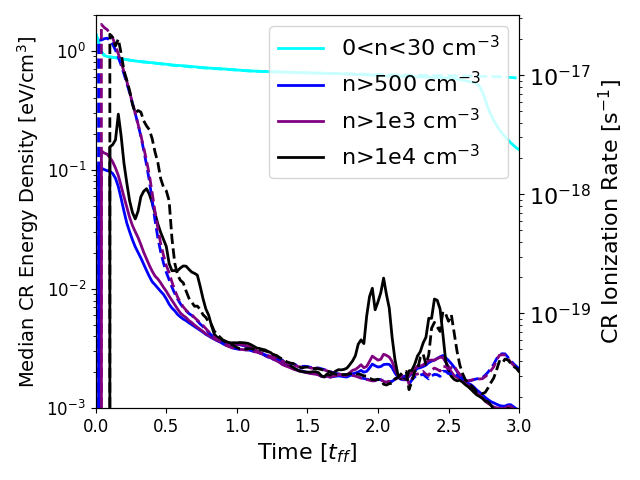}
\includegraphics[width=0.49 \textwidth]{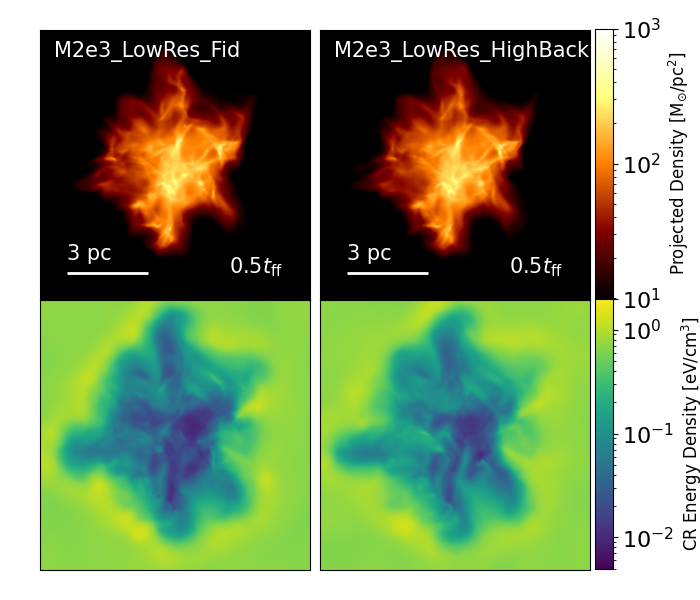}
\caption{Left: Time evolution of the median CR energy density (left axis) and the corresponding CRIR (right axis) for CRs in the indicated gas density range or lower limit for the M2e3\_LowRes\_Fid (solid) and M2e3\_LowRes\_HighBack (dashed) simulations. Right: Projected gas density and CR energy density for the M2e3\_LowRes\_Fid and M2e3\_LowRes\_HighBack simulations at $0.5 t_{\rm ff}$.}
\label{fig:m2e3_lowres}
\end{figure}

The left plot in Figure \ref{fig:m2e3_lowres} shows the evolution of the median CR energy density at different gas density cutoffs for the simulation with the fiducial setup (solid lines) and the simulation with the high initial cloud CR energy density (dashed lines). At all gas densities $>$ 500 $\rm cm^{-3}$, the results from the two simulations are indistinguishable after $\sim 0.75 t_{\rm ff}$. Although the cloud in the M2e3\_LowRes\_HighBack simulation begins with a higher initial CR density, the CRs lose energy due to streaming instability energy losses at a faster rate than in the M2e3\_LowRes\_Fid simulation. To illustrate that this does not impact the evolution of the cloud, the right panel of Figure \ref{fig:m2e3_lowres} shows the projected gas and CR energy density for both of these simulations at $0.5 t_{\rm ff}$. Although the CR energy density is lower in the interior of the cloud for the M2e3\_LowRes\_Fid simulation, the gas density configuration looks qualitatively the same as in the M2e3\_LowRes\_HighBack simulation. Since star formation does not being until $\sim 0.75 t_{\rm ff}$, we do not expect these early variations in the CR energy density to have an impact on the outcome of the simulations.

\bibliography{master}{}
\bibliographystyle{aasjournal}

\end{document}